\documentclass[aps,prx,superscriptaddress,twocolumn,floatfix]{revtex4-1}
\usepackage{amsmath}
\usepackage{amssymb}
\usepackage{amsthm}
\usepackage{graphicx}
\usepackage{graphics}
\usepackage{subfigure}
\usepackage{color}
\usepackage{bm}
\usepackage{hyperref}
\usepackage{rotating}
\def\t#1{\widetilde{#1}}

\newcommand{\ket}[1]{|#1\rangle}

\newcommand{\be}{\begin{equation} }
\newcommand{\ee}{\end{equation} }
\newcommand{\ba}{\begin{eqnarray} }
\newcommand{\ea}{\end{eqnarray} }

\newcommand{\down}{\downarrow}

\newcommand{\Z}{\mathbb{Z}}
\newcommand{\C}{\mathbb{C}}

\newcommand{\bpm}{\begin{pmatrix}}
\newcommand{\epm}{\end{pmatrix}}
\newcommand{\bmm}{\begin{matrix}}
\newcommand{\emm}{\end{matrix}}
\newcommand{\up}{\uparrow}

\begin{document}
\title{Lattice Model for Fermionic Toric Code}
\author{Zheng-Cheng Gu}
\affiliation{Perimeter Institute for Theoretical Physics, Waterloo, Ontario, N2L2Y5, Canada}
\author{Zhenghan Wang}
\affiliation{Microsoft Station Q and Department of Mathematics, University of California, Santa Barbara, CA 93106, USA}
\author{Xiao-Gang Wen}
\affiliation{Perimeter Institute for Theoretical Physics, Waterloo, Ontario, N2L2Y5, Canada}
\affiliation{Department of Physics, Massachusetts Institute of
Technology, Cambridge, Massachusetts 02139, USA}
\date{\today}

\begin{abstract}
The $Z_2$ topological order in $Z_2$ spin liquid and in exactly solvable Kitaev
toric code model is the simplest topological order for 2+1D bosonic systems.
More general 2+1D bosonic topologically ordered states can be constructed via
exactly solvable string-net models.  However, the most important topologically
ordered phases of matter are arguably the fermionic fractional quantum Hall
states.  Topological phases of matter for fermion systems are strictly richer
than their bosonic counterparts because locality has different meanings for the
two kinds of systems.  In this paper, we describe a simple fermionic version
of the toric code model to illustrate many salient features of fermionic
exactly solvable models and fermionic topologically ordered states.

\end{abstract}


\maketitle

\section{Introduction}
Kitaev's toric code model \cite{Kit03} and the
string-net models \cite{LW05,CGW1038} are exactly solvable lattice models that
produce a large class of topologically ordered phases of
matter\cite{Wtop,Wrig,WNtop} for boson/spin systems.  But rigorously solvable
lattice models for fermion systems are much less developed even though the most
important topologically ordered phases of matter are arguably the fractional
quantum Hall states\cite{TSG8259} in two dimensional fermionic systems.
On the other hand, recent discovery on the possibility of $\Z_2$ spin liquid phase in interacting fermion systems, e.g., Hubbard model on honeycomb lattice\cite{ZiyangZ2}, raises an important question that whether such a $\Z_2$ spin liquid phase has the same topological order as
toric code model or not.

In this paper, we introduce a fermionic version of the toric code model to illustrate
many salient features of rigorously solvable lattice models of topologically
ordered phases of matter for fermion systems. We show that this simple fermionic model has the same topological entanglement entropy and ground state
degeneracies on high-genus surfaces as the toric code model, nevertheless, it is described by a different topological order, characterized by its distinguishable braiding $T$ and $S$ matrices for quasiparticles. This simple fermionic model can
be generalized to a large class of exactly solvable interacting fermionic models
that we have obtained in a recent paper, which may lead to a classification of
2+1D fermionic topological orders with gapped boundaries\cite{Gu2010}. It is worthwhile to mention that
systems with topological flat bands and strong spin orbital coupling\cite{flatband1,flatband2,flatband3} could be a natural place to search such kinds of new fermionic topological orders.

Topological quantum field theories (TQFTs) are low energy effective theories for topological phases of matter for boson systems.  Similarly, spin TQFTs~\footnote{Here spin means that the space and space-time manifolds are endowed with spin structures, which are the extra data for the Dirac operator to be well-defined on general manifolds} are low energy effective theories for topological phases of matter for fermion systems.  The low energy effective TQFT for the toric code model is the Chern-Simons theory\cite{MMS0105,KLW0834}
\begin{align}
\mathcal{L} = \frac{K_{IJ}}{4\pi} \epsilon^{\lambda \mu \nu} a_{I\lambda} \partial_\mu a_{J\nu}
\end{align}
with $K$-matrix\cite{BW9045,R9002,FK9169,WZ9290,KW9327,BM0535,KS1193}
\begin{equation}
K^{TC}_{IJ} = \bpm 0 & 2 \\ 2 & 0 \epm. \label{Z2}
\end{equation}
By the fermionic toric code, we mean the topological phase of matter whose low energy effective spin TQFT is the Chern-Simons theory with $K$-matrix
\begin{equation}
K^{fTC}_{IJ} = \bpm 0 & 2 \\ 2 & 1 \epm, \label{fZ2}
\end{equation}
or its dual
\begin{equation}
\quad K^{\overline{fTC}}_{IJ} = \bpm 0 & 2 \\ 2 & 3 \epm. \label{cfZ2}
\end{equation}
The above two $K$ matrices can be regarded as the twisted version of Eq. (\ref{Z2}), and they have the same ground state degeneracies on high-genus surfaces as the toric code model due to $|\det[K^{TC}]|=|\det[K^{fTC}]|=|\det[K^{\overline{fTC}}]|$.
Nevertheless, since the diagonal elements of the above two $K$ matrices contain odd integers\cite{WZ9290,KW9327}, they can not be realized in any local bosonic system and that is why they are named as fermionic toric code.

The toric code model is a lattice realization of the $\Z_2$ gauge theory
\cite{K7959,RS9173,Wsrvb,Wen03,FNSWW04}. This model can be defined for any
lattice $\Gamma$ in any closed surface $\Sigma$.  We denote the set of sites
(or vertices), links (or edges), and plaquettes (or faces) of $\Gamma$ as
$V(\Gamma), E(\Gamma)$, and $F(\Gamma)$, respectively.  The physical degrees of
freedom are qubits living on the links (or edges) of $\Gamma$, so the local
Hilbert space $L_\Gamma=\bigotimes_{e\in E(\Gamma)}\C^2$ for the toric code is
a tensor product of $\mathbb{C}^2$ over all links of $\Gamma$.  The Hamiltonian
reads:
\begin{eqnarray*}
 H_{TC}=  -\sum_{v\in V(\Gamma)} \frac{1}{2}(1+\prod_{i\in v} \sigma^z_i)-\sum_{p\in F(\Gamma)} \frac{1}{2}(1+\prod_{i\in p} \sigma^x_i),
\end{eqnarray*}
where $\prod_{i\in p}\sigma^x_i$ is the product of the Pauli $\sigma^x_i$ over all links around
a plaquette $p$, and is called {\it the plaquette term}.  Similarly, $\prod_{i\in
v}\sigma^z_i$ is the product of the Pauli $\sigma^z_i$ over all links around a vertex $v$, and
is called {\it the vertex term}.  The ground state $\ket{\Psi_{Z_2}}$ of
$H_{TC}$ is exactly known since all terms of $H$ commute with each other.
The string-net language provides an intuitive way to understand the ground state wavefunction: if we
interpret the $\sigma^z = 1$ and $\sigma^z = -1$ states on a single link as
the absence or presence of a string, respectively, the ground state on the sphere $S^2$ is a
superpositions of all closed string states:
\begin{eqnarray}
\ket{\Psi_{TC}}=\sum_{X \rm{closed}} \ket{X}. \label{Z2wavefunction}
\end{eqnarray}
If the ground state wavefunction Eq.(\ref{Z2wavefunction}) is put on a torus, there are four different
topological sectors, characterized by the even/odd number of large strings wrapping around the torus in both directions.

The $\Z_2$-electric charge $e$ can be described as the ends of a string.  In general, elementary excitations in topological phases of matter are anyons, which
are modeled algebraically by a unitary modular category (UMC).  The UMC for the toric code is the Drinfeld double of $\Z_2$ or $SO(16)_1$, whose full data are well-known.

Now, a natural question is: Can we construct similar rigorously solvable lattice Hamiltonian for $K$ matrices Eq.(\ref{fZ2}) and Eq.(\ref{cfZ2})? This will be the main focus of this paper.
The rest of the paper is organized as follows: In Section II, we discuss the basic construction of the rigorously solvable lattice Hamiltonian on honeycomb lattice. In Section III, we describe the $\Z_2$-graded fusion rule and fermionic associativity relations, and show that the rigorously solvable lattice Hamiltonian can be consistently generated from these rules. In Section IV, we write down the explicit form of the ground state wavefunction and discuss ground state degeneracy on torus. In Section V, we compute the braiding $T$ and $S$ matrices of the anyon excitations of the rigorously solvable lattice Hamiltonian. Finally, there is a short discussion for the general mathematic structure behind this model.

\section{Fermionic local Hilbert spaces and exactly solvable lattice Hamiltonians}
\subsection{Fermionic local Hilbert spaces}
\begin{figure}[t]
\begin{center}
\includegraphics[width=8cm]{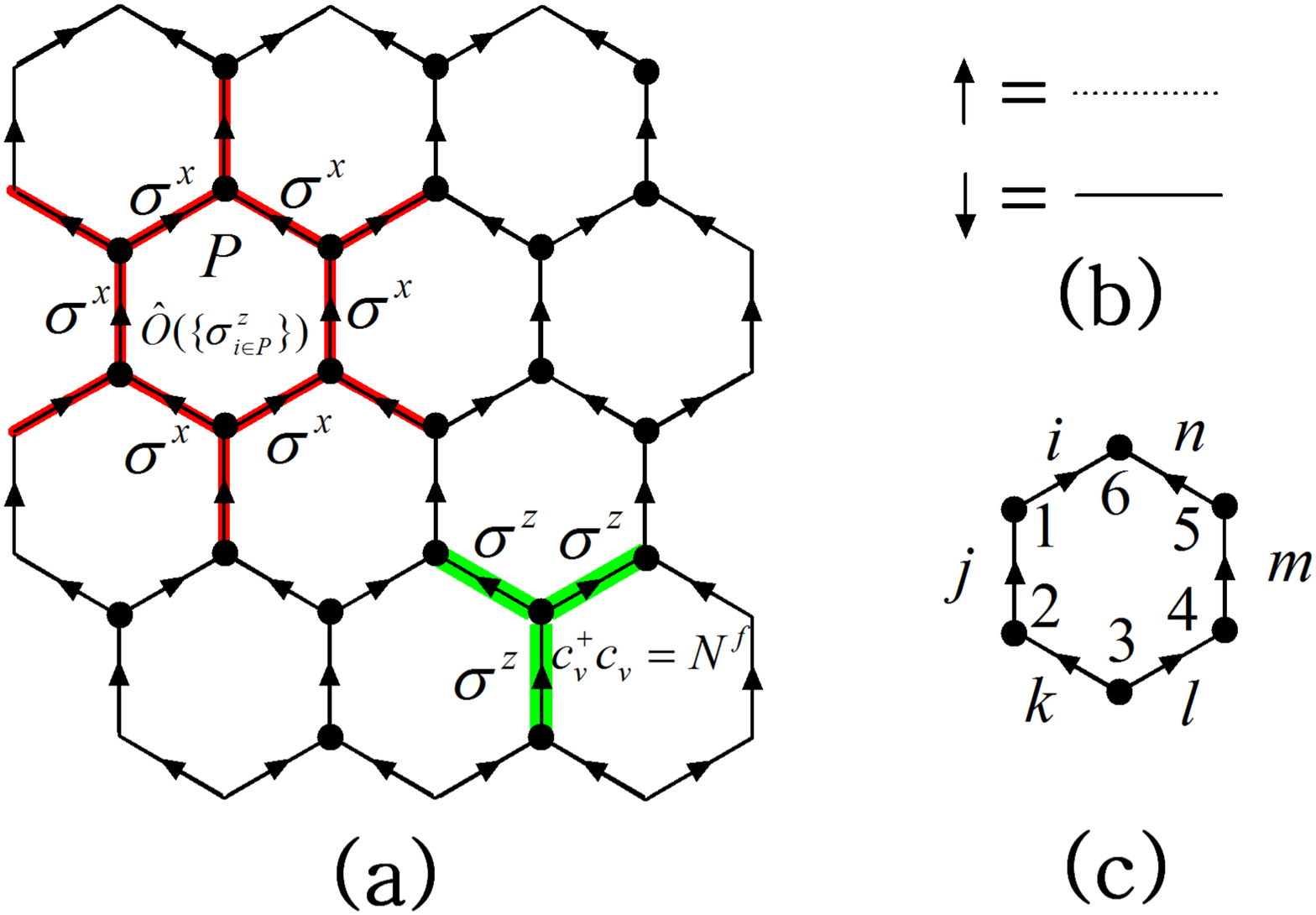}
\vskip -0.5cm
\caption{The honeycomb lattice has a fixed branching structure (branching arrows from down to up). (a) A vertex term is not just a product of $\sigma^z$ surrounding the vertex, but also contains a projector
for the fermion number on the vertex, which is determined by the so-called $\Z_2$-graded fusion rule. A plaquette term only acts on the subspace of the projector $P_p=\prod_{v\in p}Q_v$, and in additional to the $\sigma^x$ surrounding the plaquette, a term $\hat O(\{\sigma_{b\in p}^z\})$ consisting of a product of fermion creation/annihilation operators is also needed. (b) We map a spin-$\up$ state to the absence of a string and a spin-$\down$ state to the presence of a string. (c) To define the operator $\hat O(\{\sigma_{b\in p}^z\})$, we label the links of a plaquette by $i,j,k,l,m,n$ and the vertices of a plaquette by $1,2,3,4,5,6$. In Table.\ref{Hamiltonian}, we compute the explicit expression of $\hat O(\{\sigma_{b\in p}^z\})$ using the fermionic associativity relations.
}\label{H}
\end{center}
\end{figure}

To define a femionic model on a lattice $\Gamma$ in a general surface $\Sigma$, we need several structures on $\Gamma$ in $\Sigma$.  One such structure is a branching structure on a trivalent lattice $\Gamma$: a choice of an arrow on each link of $\Gamma$ so that around each site (or vertex), the three arrows are never all-in or all-out.  Conceptually, a branching structure is amount to a choice of a local coordinate system around each site.  One obvious branching is a global ordering of all vertices, and then an arrow on any link goes from the smaller vertex to the bigger one in this ordering.  For the honeycomb lattice below, we choose the branching that all arrows go upwards in the plane as shown in Fig.\ref{H}.  Another additional structure on $\Gamma$ in $\Sigma$ is a lattice version of a spin structure of an oriented surface.  As usual, we will encode the spins structures as boundary conditions here.

Each site of the lattice $\Gamma$ can be either occupied or not by a spinless fermion, so the fermion occupation number $N^f$  of a site $v$ is either $0$ or $1$.  Let $|0_V\rangle$ be the ground state of no fermions on all sites. Then a generating set of the Fock space is given by $\prod _{v \in I}c^\dagger_v|0_V\rangle$, where $I\subset V$ is a subset of all sites $V$ including the empty set.  The full local Hilbert space for our model is
$$L^{fTC}_\Gamma=\bigoplus_{I\subset V} (\prod _{v \in I}c^\dagger_v|0_V\rangle \bigotimes (\otimes_{e\in E(\Gamma)} \mathbb{C}^2)).$$
As a vector space, the fermionic Hilbert space is the same as the tensor product $\bigotimes_{v\in V(\Gamma)}\C^2 \otimes \bigotimes_{e\in E(\Gamma)}\C^2.$  But the Fock space structure over the sites means that a local Hamiltonian for a fermion system is non-local when regarded as one for a boson system.

\begin{table*}[t]
 \centering
 \begin{tabular}{ |c|c|c|c|}
\hline
$ \sigma_i^z,\sigma_j^z,\sigma_k^z,\sigma_l^z,\sigma_m^z,\sigma_n^z $ & $\hat O (\sigma_i^z,\sigma_j^z,\sigma_k^z,\sigma_l^z,\sigma_m^z,\sigma_n^z)$
&
$ \sigma_i^z,\sigma_j^z,\sigma_k^z,\sigma_l^z,\sigma_m^z,\sigma_n^z$ & $\hat O (\sigma_i^z,\sigma_j^z,\sigma_k^z,\sigma_l^z,\sigma_m^z,\sigma_n^z)$
\\
\hline
111111 & $\pm ic_6^\dagger c_3^\dagger$ &
&  \\
\hline
-111111 & $c_3^\dagger c_1$ &
1-11111 & $c_6^\dagger c_3^\dagger c_2^\dagger c_1^\dagger $ \\
11-1111 & $c_6^\dagger c_2$ &
111-111 & $c_6^\dagger c_4$ \\
1111-11 & $-c_6^\dagger c_5^\dagger c_4^\dagger c_3^\dagger $ &
11111-1 & $-c_5 c_3^\dagger$ \\
\hline
-1-11111 & $c_3^\dagger c_2^\dagger$ &
1-1-1111 & $c_6^\dagger c_1^\dagger $ \\
11-1-111 & $-c_6^\dagger c_4 c_3 c_2$ &
111-1-11 & $-c_6^\dagger c_5^\dagger$ \\
1111-1-1 & $c_4^\dagger c_3^\dagger $ &
-11111-1 & $c_6 c_5 c_3^\dagger c_1$ \\
\hline
-11-1111 & $\mp ic_2 c_1$ &
1-11-111 & $\mp ic_6^\dagger c_4 c_2^\dagger c_1^\dagger $ \\
11-11-11 & $\pm ic_6^\dagger c_5^\dagger c_4^\dagger c_2$ &
111-11-1 & $\pm ic_5c_4$ \\
-1111-11 & $\pm ic_5^\dagger c_4^\dagger c_3^\dagger c_1 $ &
1-1111-1 & $\pm ic_5 c_3^\dagger c_2^\dagger c_1^\dagger$ \\
\hline
-111-111 & $\mp ic_4 c_1$ &
1-111-11 & $\pm ic_6^\dagger c_5^\dagger c_4^\dagger c_3^\dagger c_2^\dagger c_1^\dagger $ \\
11-111-1 & $\pm ic_5 c_2$ & &\\
\hline
111-1-1-1 & $1$ &
11-1-1-11 & $c_6^\dagger c_5^\dagger c_3 c_2 $ \\
1-1-1-111 & $-c_6^\dagger c_4 c_3 c_1^\dagger$ & &\\
\hline
-11-1-111 & $\pm ic_4 c_3c_2 c_1$ &
1-11-1-11 & $\pm ic_6^\dagger c_5^\dagger c_2^\dagger c_1^\dagger $ \\
11-11-1-1 & $\mp ic_4^\dagger c_2$ &
-111-11-1 & $\mp ic_6 c_5c_4 c_1$ \\
-1-111-11 & $\pm ic_5^\dagger c_4^\dagger c_3^\dagger c_2^\dagger $ &
1-1-111-1 & $\pm ic_5 c_1^\dagger$ \\
\hline
1-11-11-1 & $c_5^\dagger c_4^\dagger c_2 c_1$ &
&  \\
\hline
\end{tabular}
 \caption{The $32$ independent nonzero cases of $\hat O(\{\sigma_{b\in p}^z\})$. The $\pm$ sign corresponds to the theory $K^{\overline{fTC}}$/$K^{{fTC}}$. }
\label{Hamiltonian}
\end{table*}

\subsection{Exactly solvable lattice Hamiltonians}
Similar to the toric code model, the lattice Hamiltonian for the fermionic toric code is expressed as a sum of commuting projectors on the honeycomb lattice $\Gamma$ with a fixed branching structure, as seen in Fig \ref{H}, and thus exactly solvable:
\begin{equation}
H_{fTC}=-\sum_{v\in V(\Gamma)}Q_v-\sum_{p\in E(\Gamma)}Q_p. \label{Hftc}
\end{equation}
Both the vertex term $Q_v$ and the plaquette term $Q_p$ are projectors defined by:
\begin{equation}
Q_v=\frac{1}{2}\left(1+\prod_{i\in v}\sigma^z_i\right)\left\{1-\left[c_v^\dagger c_v-{N^f(\{\sigma_{a\in v}^z\})}\right]^2\right \}
\end{equation}
and
\begin{equation}
Q_p=\frac{1}{2}\left(1+\hat O(\{\sigma_{b\in p}^z\}) \prod_{i\in p}\sigma^x_i\right)P_p.
\end{equation}

Similar to the toric code model, here we also map the spin-$\up$ or spin-$\down$ state to the absence or presence of a string.
The projector $Q_v$ is the same as the closed string constraint for toric code, except that fermions are decorated onto the closed string
according to the fermion occupation number function $N^f(\{\sigma_{a\in v}^z\})=$ $0$ or $1$, and such a decoration is consistently assigned by the so-called $\Z_2$-graded fusion rule below. The term $P_p=\prod_{v\in p}Q_v$ is also a projector, and $\hat O(\{\sigma_{b\in p}^z\})$ consists of a product of fermion creation/annihilation operators and a phase factor, which can be computed from the fermionic associativity relations below. We note that the product of fermion creation/annihilation operators is to make sure that the initial and final closed string configurations are all decorated by fermions in a consistent way.

\begin{figure}[t]
\begin{center}
\includegraphics[width=6cm]{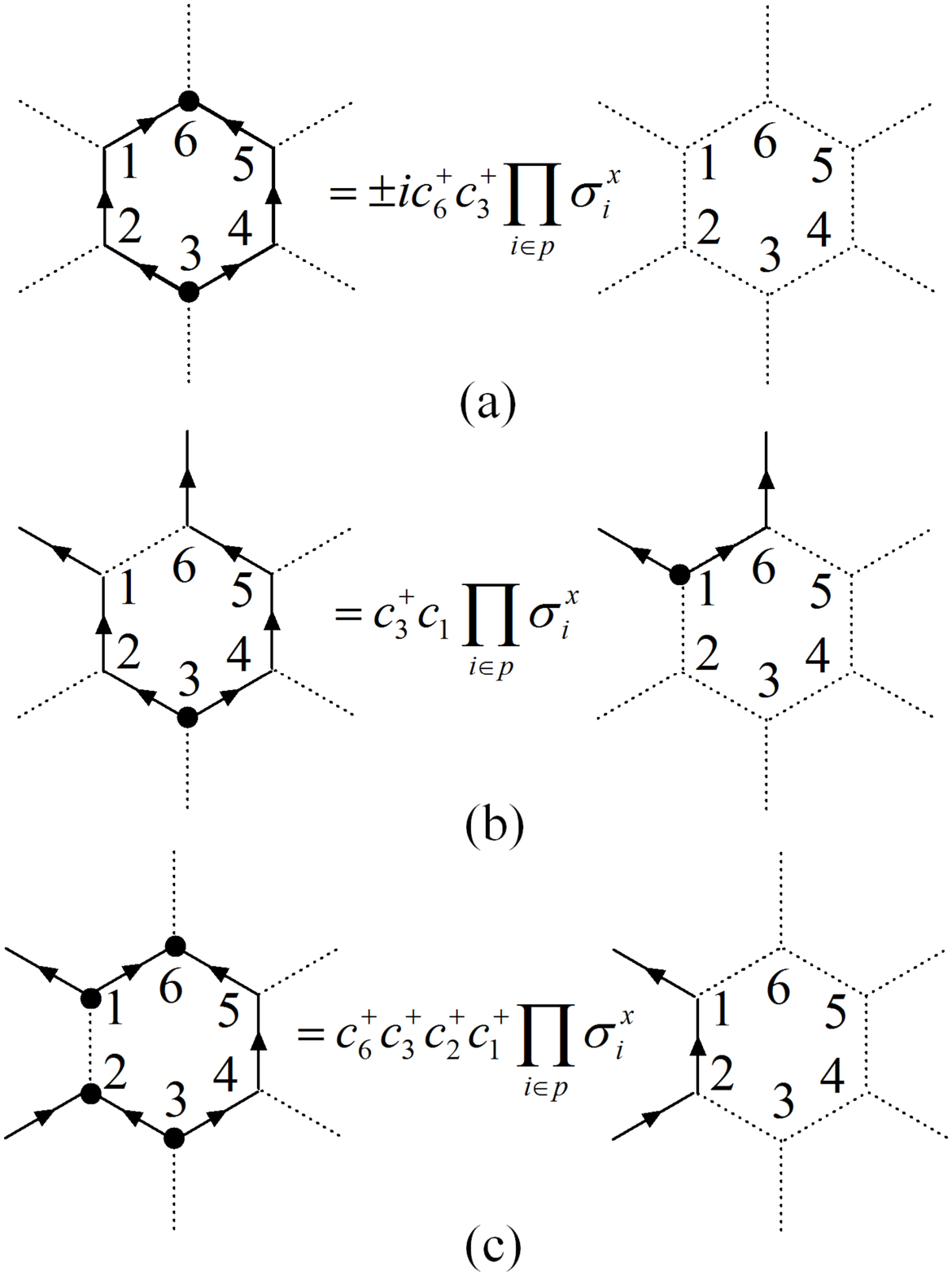}
\caption{Examples of the $Q_p$ operator acts on closed string configurations for a hexagon. (a), (b) and (c) correspond to the first three terms in table \ref{Hamiltonian}.
We use solid/dot lines to represent presence/absence of strings and black dots to represent the occupation of fermions}\label{Hexample}
\end{center}
\end{figure}

Since $Q_p$ is a Hermitian operator, it is easy to see that $\hat O^\dagger(\{\sigma_{b\in p}^z\})=\hat O(\{-\sigma_{b\in p}^z\})$. In Table.\ref{Hamiltonian}, we list the $32$ independent cases of $\hat O(\{\sigma_{b\in p}^z\})$, and the  other $32$ cases follow from $\hat O^\dagger(\{\sigma_{b\in p}^z\})=\hat O(\{-\sigma_{b\in p}^z\})$. In Fig.\ref{Hexample}, we plot the first three terms in Table.\ref{Hamiltonian},
where we use solid/dot lines to represent presence/absence of strings and black dots to present the occupation of fermions.

It is not hard to check that all $Q_v$ commute with themselves and all $Q_p$. The self-consistency of the fermionic associativity relations below will imply that all the $Q_p$ also commute with themselves. Once we set up our theory, this follows from the fact that the actions of $Q_pQ_{p^\prime}$ and $Q_{p^\prime}Q_p$ on any closed loop configuration lead to the same final configuration. In the following, we explain how to construct the $\Z_2$-graded fusion rule and the fermionic associativity relations.

\section{$\Z_2$-graded fusion rules and fermionic associativity relations}
\subsection{Fusion rules and associativity relations -- the algebraic approach to topological states}
\begin{figure}[h]
\begin{center}
\vskip -1.5cm
\includegraphics[width=6cm]{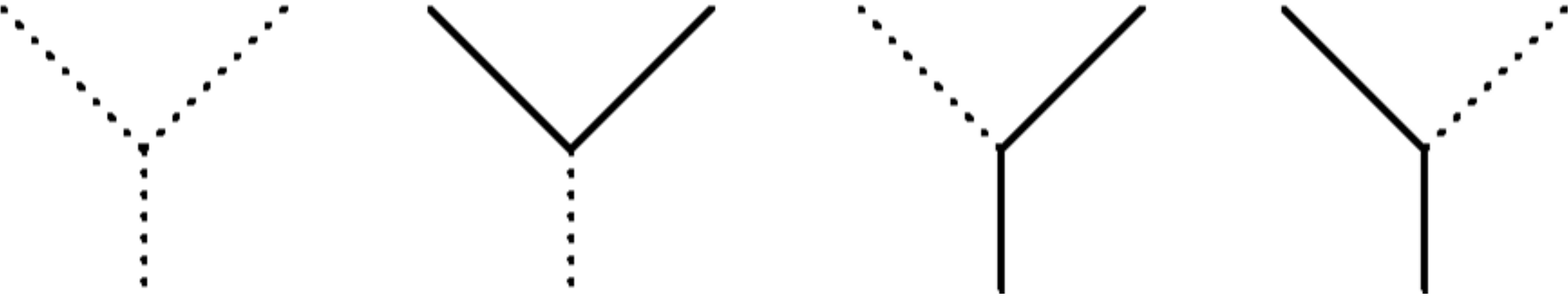}
\vskip -1.5cm
\caption{The fusion rules for the toric code model. Here we use solid/dot lines to represent presence/absence of ($\Z_2$) strings.
}\label{fusion}
\end{center}
\end{figure}

To illustrate the basic idea of how to construct the above exactly solvable lattice models and compute their basic topological properties, e.g., ground state degeneracy on torus, braiding statistics of quasi-particle excitations,
we begin with an algebraic way of constructing the toric code model.
The key idea is to introduce the fusion rules and associativity relations for its ground state wavefunction(a fixed point wavefunction from the renormalization group point of view). Indeed, the toric code model on a trivalent lattice is the same as the string-net model with the $\Z_2$ unitary fusion category (UFC) as input.
Without loss of generality, for any string-net model on a trivalent lattice, we can use a function $N_{ijk}=N_{jki}=N_{kij}=0,1$\cite{LW05,CGW1038} to characterize the fusion rule, and use $N_{ijk}=1$ to represent admissible configurations in the ground state wavefunction of the string-net model, where $i,j,k$ label string types and $i,j,k=0$ labels the string vacuum.  The $\Z_2$ fusion rule is defined by $N_{110}=N_{101}=N_{011}=N_{000}=1$(Here $i,j,k=0,1$ corresponds to absence and presence of a $\Z_2$ string) and $N_{ijk}=0$ otherwise, as seen in Fig.\ref{fusion}.

\begin{figure}[bt]
\begin{center}
\includegraphics[width=8cm]{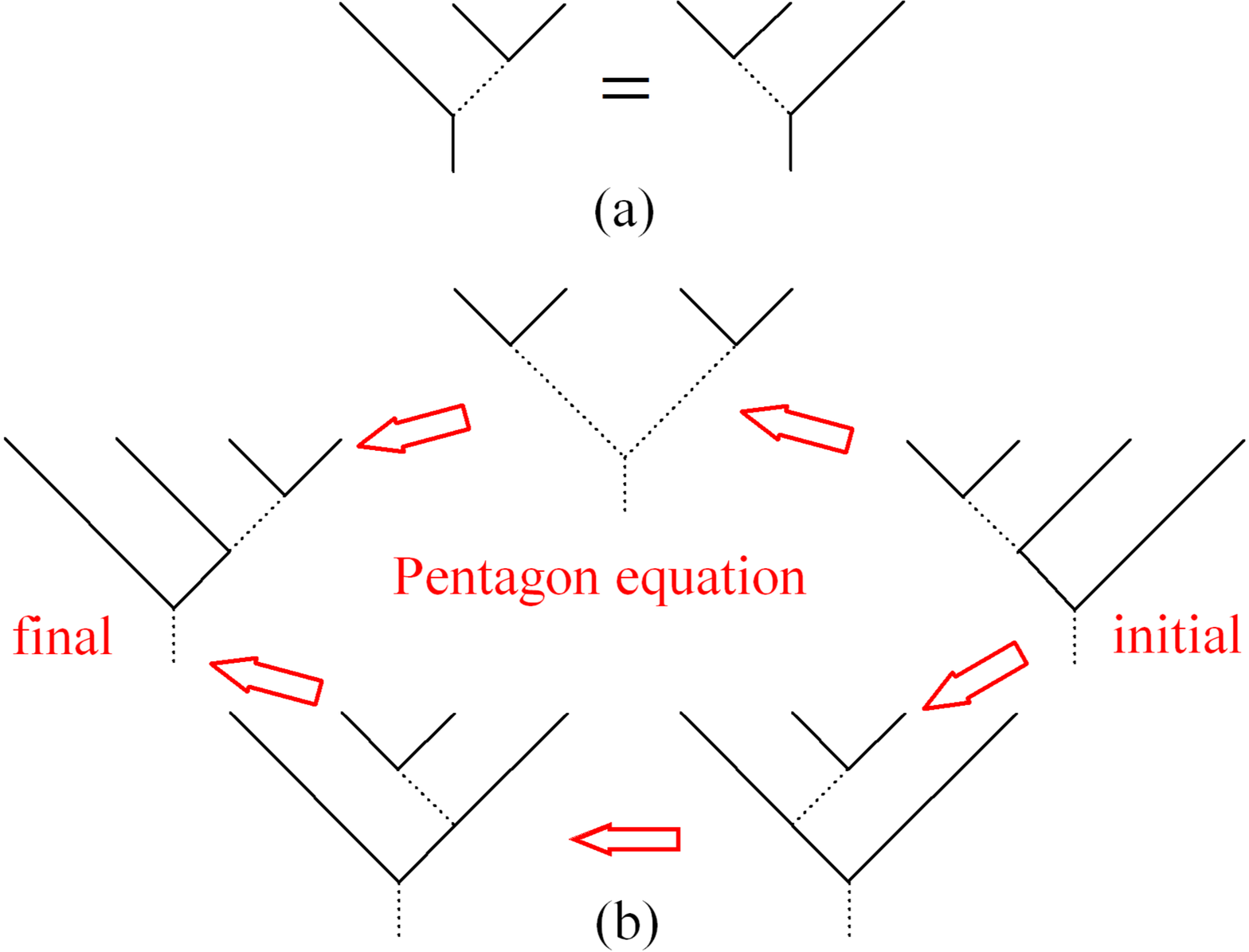}
\caption{(Color online)(a) The only nontrivial associativity relation.
(b) An example of pentagon equation that issues the self-consistency of the associativity relations for toric code model.
}\label{Pentagon}
\end{center}
\end{figure}

Physically, the associativity relations are local rules that allow us to compute the amplitudes of admissible configurations for the ground state wave functions.
Fig.\ref{Pentagon}(a) illustrates the only nontrivial associativity relation for the toric code model.(Other associativity relations just trivially change the shape or length of a string.) In terms of UMC language, the associativity relations are called as the $6j$ symbols, which are inputs of the underlying $\Z_2$ UFC and satisfy the well-known Pentagon equations.

Let us explain a little bit more about the physical meaning of the Pentagon equations. As seen in Fig.\ref{Pentagon}(b),
we assume that the initial(reference) and final configurations only differ in a local region. Starting from the initial(reference) configuration whose amplitude is known in the ground state wavefunction, we can compute the amplitude of the final configuration by applying the associativity relations. However, we have two different paths(upper and lower paths) to do so, and the Pentagon equations issue that the amplitude of the final configuration is independent on particular choice of paths.

Furthermore, the consistent associativity relations will also allow us to construct an exactly solvable Hamiltonian consisting of commuting projectors with a constant energy gap and finite ground state degeneracy on the torus. Apparently, the vertex terms in $H_{TC}$ come from the $\Z_2$ fusion rule and commute with each other, and the plaquette terms in $H_{TC}$ can be computed from associativity relations, which only act within the closed string subspace. The self-consistency of associativity relations implies that plaquette terms commute with each other as well. In the following, we will generalize these basic concepts into interacting fermion systems and discuss the simplest example -- fermionic toric code.

\subsection{$\Z_2$-graded fusion rules}
\begin{figure}[h]
\begin{center}
\vskip -1.5cm
\includegraphics[width=7cm]{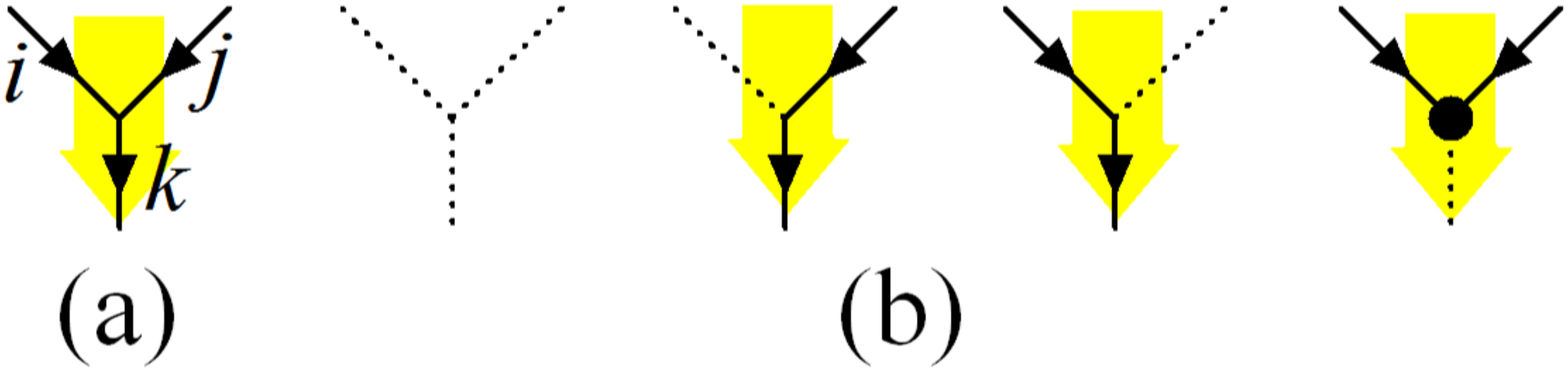}
\vskip -1.5cm
\caption{(Color online) (a) The branched fusion. The yellow arrow indicates the local direction that induces a branching structure for the links connecting to a vertex. (b) The $\Z_2$ fusion rules with a graded structure.  The solid dot represents a fermion.
}\label{branchfusion}
\end{center}
\end{figure}

Our input for the fermionic toric code will be a $\Z_2$-graded version of the $\Z_2$ UFC.  As in the toric code case, the input category has two particle types denoted as $0,1$ with the only non-trivial fusion rule $1\otimes 1=0$.  The fusion space for a fermionic category is a $\Z_2$-graded Hilbert space.  Therefore, the fusion coefficient $N_{ij}^k$ is a sum of $N_{ij}^{k,b}$ and $N_{ij}^{k,f}$ for the dimensions of the even and odd parts of the $\Z_2$-graded Hilbert space.
The branching structure removes the cyclic symmetry in the indices of $N_{ijk}$. As shown in Fig.\ref{branchfusion} (a), a branching structure is induced by a local direction. We use a new notation $N_{ij}^k$ with lower indices representing incoming arrows and upper indices representing outgoing arrows. So we have:
\begin{equation}
N_{ij}^k={N}_{ij}^{k,b}+{N}_{ij}^{k,f}
\end{equation}
with ${N}_{ij}^{k,b},{N}_{ij}^{k,f}=0$ or $1$. Here we also assume ${N}_{ij}^{k,b}={N}^{ij,b}_{k}$ and ${N}_{ij}^{k,f}={N}^{ij,f}_k$. For a fermion parity odd fusion channel, a fermion will be present at the corresponding vertex and we use a solid dot to represent the fermion. To construct the fermionic toric code model, we assign ${N}_{00}^{0,b}={N}_{01}^{1,b}={N}_{10}^{1,b}=1$, ${N}_{11}^{0,f}=1$, as shown in Fig.\ref{branchfusion} (b), and all the others ${N}_{ij}^{k,b}={N}_{ij}^{k,f}=0$. The reason why the above graded structure for $N_{ij}^k$ works is the following. We note that ${N}_{00}^{0,b}=1$ represents the vacuum configuration, thus it is always admissible, and with an even fermion parity for any local theory. ${N}_{01}^{1,b}={N}_{10}^{1,b}=1$ is due to the fact that the above fusion rules only make the length of strings longer or shorter, and as a topological theory, the fermion parity of a wavefunction should not depend on its length. Finally, ${N}_{11}^{0,f}=1$ is possible since the string reverses its branching arrow when it goes through the vertex.

Therefore, in the fermionic toric code, each fusion coefficient $N_{ij}^k$ is either fermion parity even or odd with only one fermion parity odd fusion space $N_{11}^{0,f}=1$.
The corresponding fermion number function $N^f(\sigma_i^z,\sigma_j^z,\sigma_k^z)$ reads:
\begin{equation}
N^f(1,1,0)=1 ;\quad N^f(\sigma_i^z,\sigma_j^z,\sigma_k^z)=0 \quad \rm{otherwise}
\end{equation}

\subsection{Fermionic associativity relations and fermionic pentagon equations}
The Hamiltonian and wave functions in our models depend on the $6j$ symbols of the input category, which are the solutions to the \emph{fermionic pentagon equations} below.  The fermionic pentagon equations guarantee mathematically the consistency of \emph{fermionic associativity relations} in fusing anyons.

\begin{figure}[bt]
\begin{center}
\includegraphics[width=8cm]{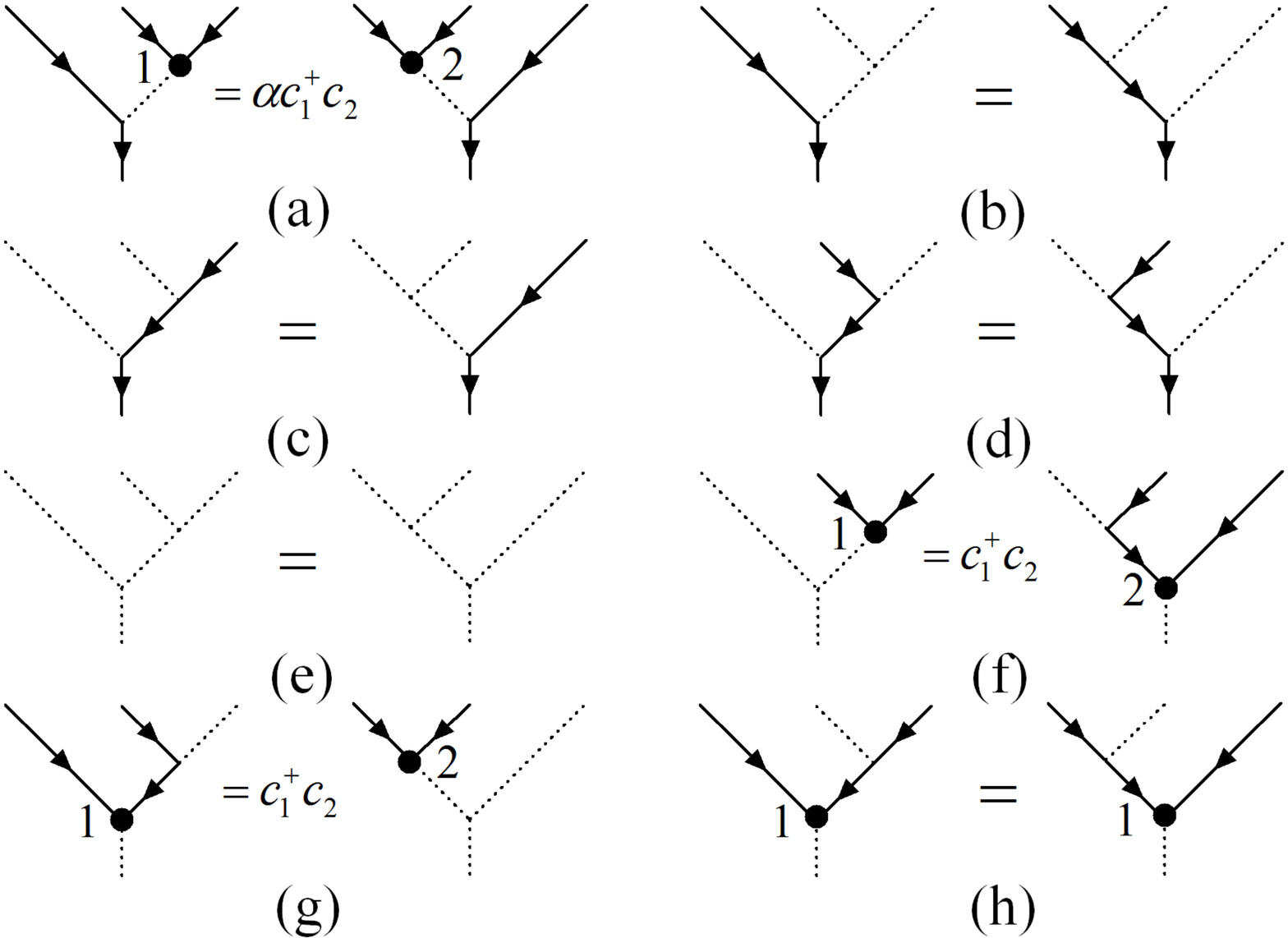}
\caption{All possible fermionic associativity relations with a fixed choice of a branching structure (branching arrows from up to down).  Only the first associativity relation is nontrivial, and all the others only change the shape/length of a string or the position of a fermion.   Associativity relations with different branching structures can be computed from the above basic relations and the bending moves.
}\label{fermionfusion}
\end{center}
\end{figure}

In Fig.\ref{fermionfusion}, we list all the possible fermionic associativity relations with the fixed branching structure (branching arrows from up to down). We note that only the first associativity relation is nontrivial and all the others only change the shape/length of a string or position of a fermion, and will not change the amplitudes of admissible configurations in the fixed point wavefunction.

\begin{figure}[bt]
\begin{center}
\vskip -1.0cm
\includegraphics[width=8cm]{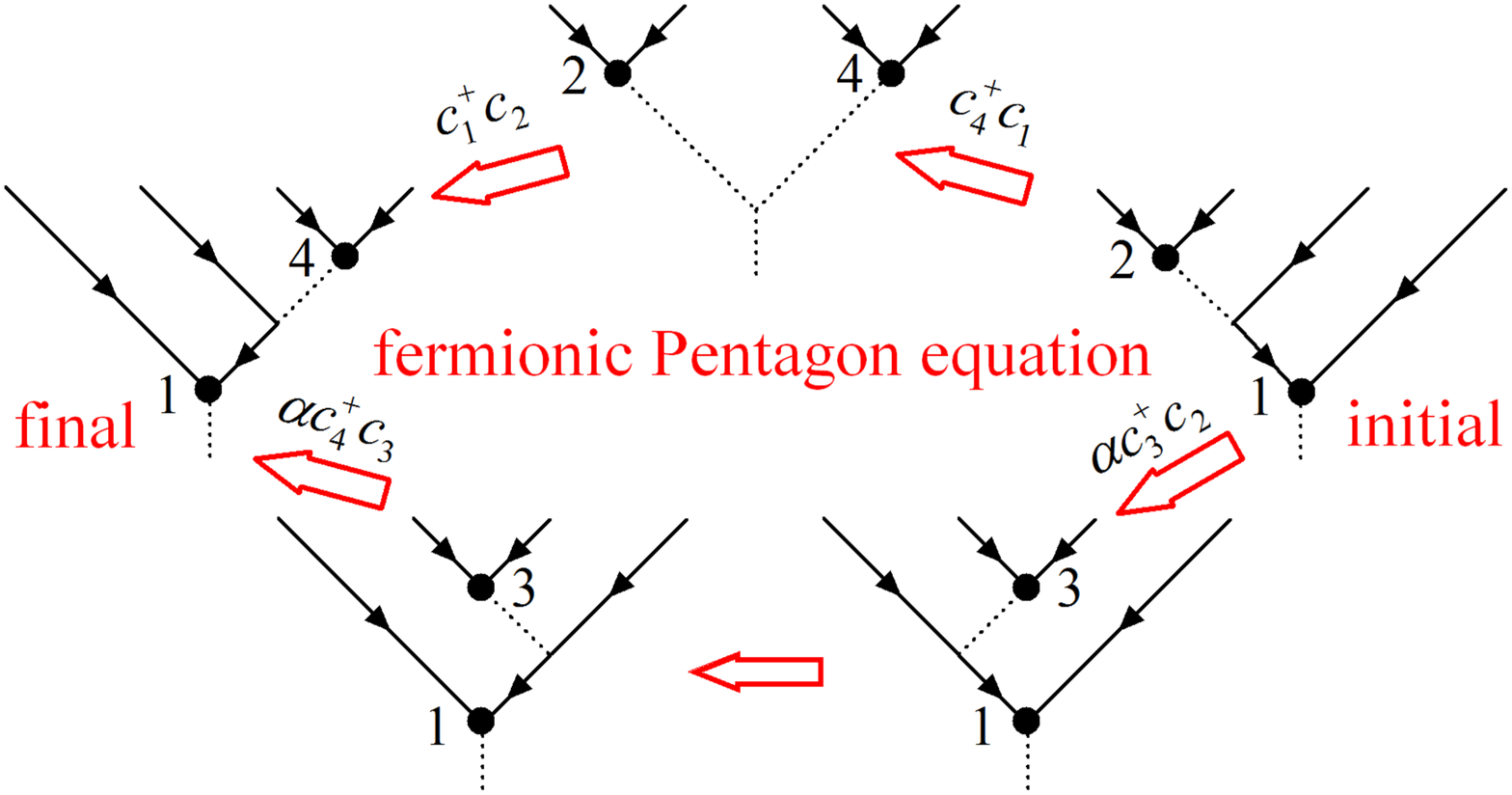}
\vskip -1.0cm
\caption{(color online)
One of the fermionic pentagon equations that the fermionic associativity relations need to satisfy.
}\label{fPentagon}
\end{center}
\end{figure}

Now we discuss our key technical result in this paper: the \emph{fermionic pentagon equations} that the fermionic associativity relations must satisfy.  The fermionic pentagon equations also have a graded structure due to the fermion sign.
Such a $\Z_2$-graded structure allows us to construct new topological phases that can not be realized in any bosonic system.
In Fig.\ref{fermionPentagon} in the supplemental materials, we list all the nontrivial fermionic pentagon relations that the fermionic associativity relations must satisfy. However, only the one of these relations in Fig.\ref{fPentagon} or Fig.\ref{fermionPentagon} (a) gives rise to a constraint on the numerical value of $\alpha$.

As shown in Fig.\ref{fPentagon}, for the fermionic toric code model, we can compute the amplitude of an admissible configuration by applying the fermionic associativity relations. However, we can choose different paths and they must give rise to the same result. If we choose the upper path, we have $|{\rm final}\rangle=c_1^\dagger c_2 c_4^\dagger c_1|{\rm initial}\rangle=-c_4^\dagger c_2|{\rm initial}\rangle$. Here $|{\rm final}\rangle$ and $|{\rm initial}\rangle$ are admissible configurations of the fixed point wavefunction that only differ in a local region (the configurations in the outside region that does not appear in Fig.\ref{fPentagon} are the same). On the other hand, the lower path implies $|{\rm final}\rangle=(\alpha c_4^\dagger c_3)(\alpha c_3^\dagger c_2)|{\rm initial}\rangle=\alpha^2 c_4^\dagger c_2|{\rm initial}\rangle$. As a result, we obtain $\alpha=\pm i$.
We note that the emergence of such a nontrivial minus sign here is because that the two different paths differ with a fermion loop.

\subsection{Bending moves}
However, there are additional complications due to the branching structure here, as for a fixed point wavefunction, we should be able to define it on trivalent lattices with arbitrary branching structures.
\begin{figure}[tb]
\begin{center}
\vskip -0.5cm
\includegraphics[width=8cm]{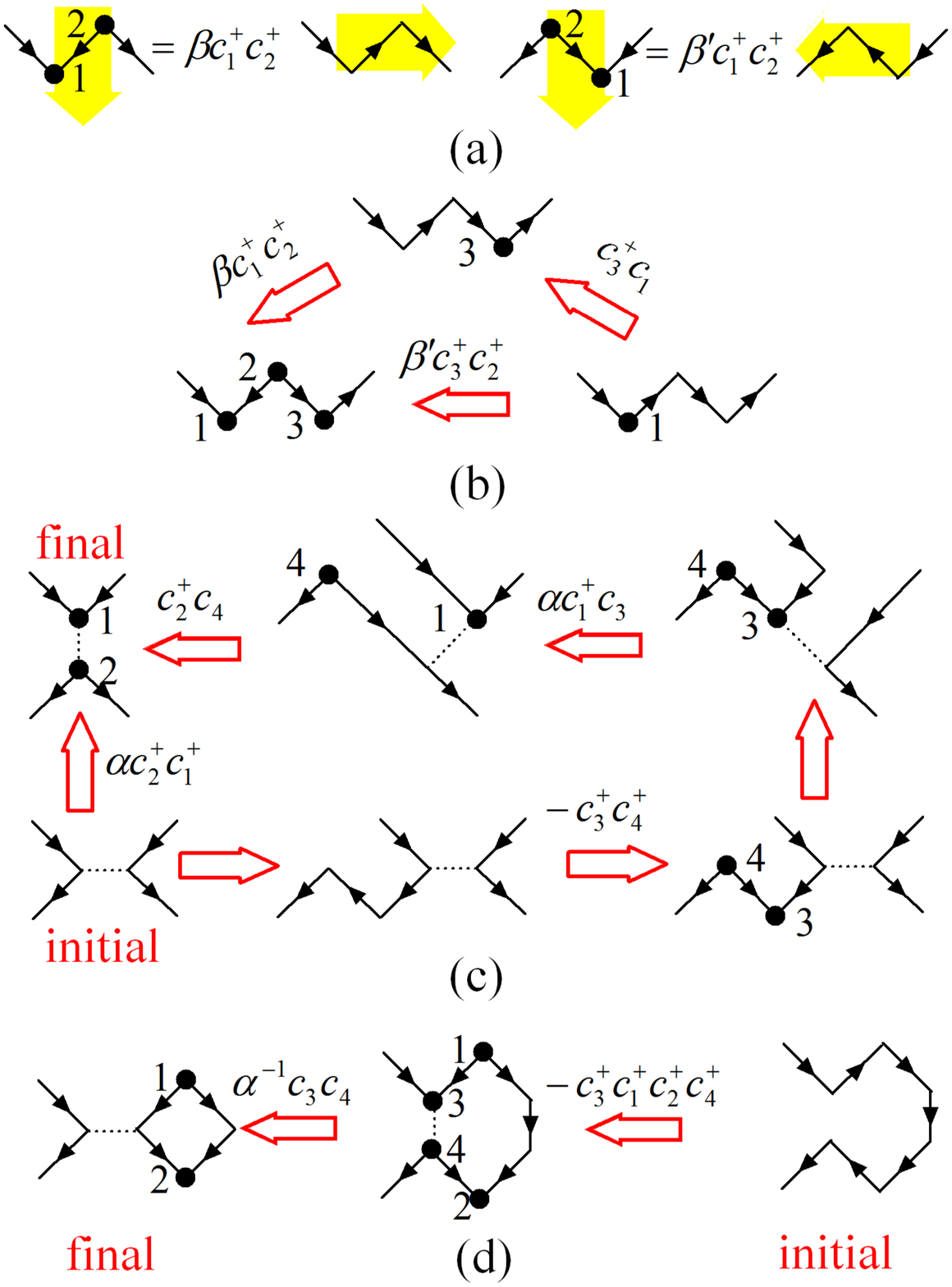}
\vskip -0.5cm
\caption{(color online)(a) The left/right handed bending moves. (b) By applying the bending moves in the upper path and lower path, we have $\beta=-\beta^\prime$. (c) Fermionic associativity relations with different branching structure can be computed through bending moves. (d) The \lq\lq loop weight" can be computed by using associativity relations.
}\label{bending}
\end{center}
\end{figure}

Indeed, by using some special associativity relations --- the \emph{bending moves}, we can generate all the branched associativity relations in a self-consistent and systematic way. The bending moves will allow us to change the branching arrow of a string locally. There are two different bending moves, and we call them left and right hand bending moves, as seen in Fig.\ref{bending} (a). We note that the left/right bending move will change the branching arrow clockwise/anti-clockwise by $90$ degrees. On the other hand, by comparing the two different associativity relations in Fig.\ref{bending} (b) from upper path and lower path, we conclude that $\beta^\prime=-\beta$, and for a unitary theory, we further have $|\beta|=1$. The actual value of $\beta$ can be chosen freely, since we can redefine $\beta$ as $\t \beta=e^{-2i\theta}\beta$ by applying gauge transformations for all the fermion creation operators, $\t c^\dagger=e^{i\theta} c^\dagger$. Meanwhile, it is important that under such a gauge transformation, $\alpha$ remains unchanged. Let us choose $\beta=1$ throughout the whole paper for convenience.

The bending moves can help us to compute associativity relations with different branching structures. For example, in Fig.\ref{bending}(c), the nontrivial associativity relation has a branching structure with two incoming and two outgoing arrows(different from the case in Fig.\ref{fermionfusion} (a) with three incoming and one outgoing arrows). We have $|{\rm final}\rangle=-\alpha c_2^\dagger c_4 c_1^\dagger c_3 c_3^\dagger c_4^\dagger|{\rm initial}\rangle=\alpha c_2^\dagger c_1^\dagger|{\rm initial}\rangle$. In Fig.\ref{bending}(d), by applying the result of Fig.\ref{bending}(c), we have $|{\rm final}\rangle=-\alpha^{-1} c_3c_4 c_3^\dagger c_1^\dagger c_2^\dagger c_4^\dagger|{\rm initial}\rangle=\alpha^{-1}c_1^\dagger c_2^\dagger|{\rm initial}\rangle$, which implies that the \lq\lq loop weight" is $\alpha^{-1}c_1^\dagger c_2^\dagger$.
Now we can compute all the fermionic associativity relations with arbitrary branching structures.

\begin{figure}[tb]
\begin{center}
\vskip -2.5cm
\includegraphics[width=9cm]{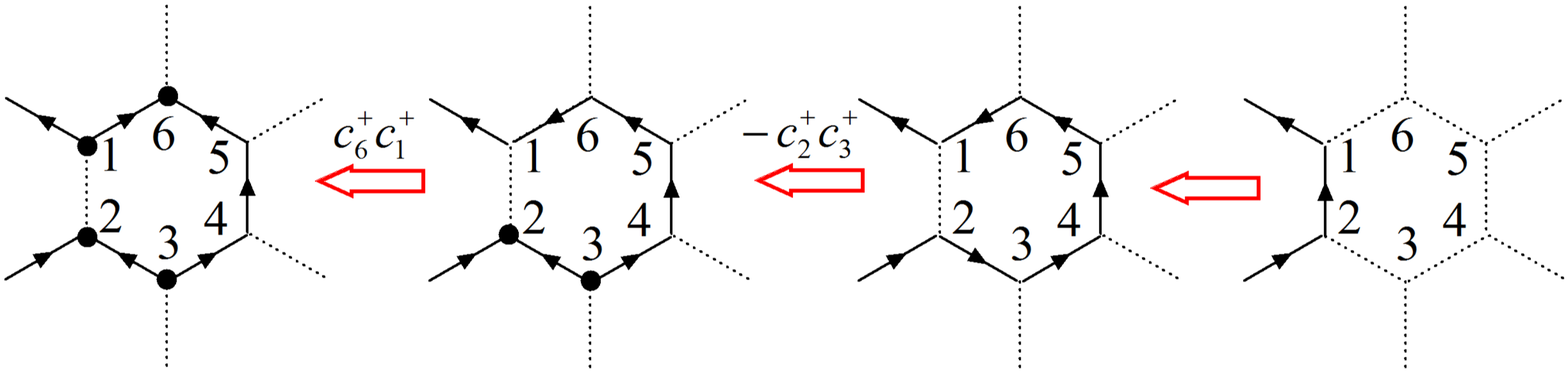}
\vskip -2.5cm
\caption{(color online)An example of evaluating the operator $\hat O(\{\sigma_{b\in p}^z\})$ for a given initial decorated closed string configuration.} \label{Hevaluate}
\end{center}
\end{figure}

\subsection{Constructing exactly solvable Hamiltonians}
To this end, we are able to explain how to construct exactly solvable Hamiltonian Eq.(\ref{Hftc}). Apparently, the projectors $Q_v$ trivially follow from the $\Z_2$-graded fusion rules, and the low energy subspace is consisting of closed strings decorated by femions on those vertices where branching arrows get reversed. The projectors $Q_P$ that act within the decorated closed strings subspace(hence all $Q_P$ commute with all $Q_v$) can be evaluated from the fermionic associativity relations with a fixed branching arrow introduced in Fig.\ref{fermionfusion}  as well as the bending moves introduced in Fig.\ref{bending}(a).

For example, for the term in Fig.\ref{Hexample} (a), the operator $\hat O(1,1,1,1,1,1)$ is nothing but the "loop weight" that have been computed in Fig.\ref{bending}(d). We note that according to Fig.\ref{bending}(d), the "loop weight" in Fig.\ref{Hexample} (a) should be $\alpha^{-1}c_3^\dagger c_6^\dagger=-\alpha^{-1}c_6^\dagger c_3^\dagger=\pm i c_6^\dagger c_3^\dagger$.
In Fig.\ref{Hexample} (b), the operator $\hat O(-1,1,1,1,1,1)$ just trivially changes the length(shapes) of a string and moves the decorated fermion to a new vertex where the string reverses its branching arrow. Therefore, it is just $c_3^\dagger c_1$ without any additional phase factor. In Fig.\ref{Hevaluate}, we show how to evaluate the operator $\hat O(\{\sigma_{b\in p}^z\})$ in Fig.\ref{Hexample} (c), where we first apply two steps of bending moves, and then apply a trivial fermonic associativity relation which just changes the length of a string. The bending moves give rise to $\hat O(1,-1,1,1,1,1)=(c_6^\dagger c_1^\dagger)(-c_2^\dagger c_3^\dagger)=c_6^\dagger c_3^\dagger c_2^\dagger c_1^\dagger$.

All the listed $\hat O(\{\sigma_{b\in p}^z\})$ terms with different initial spin configurations in Table.\ref{Hamiltonian} can be evaluated in similar ways, and the fermionic Pentagon equations will guarantee that all the $Q_P$ operators commute with each other.

\section{Ground state wave functions and degeneracy on torus}
Using the fermionic associativity relations, we obtain the ground state wavefunction for the fermionic toric code model on the sphere as follows:
\begin{equation}
\ket{\Psi_{\text{fTC}}}=\sum_{X \rm{closed}}
\sigma(X)\alpha^{n(X)}\prod_{v\in X} {(c_v^\dagger)}^{N^f(\{\sigma_{a\in v}^z\})} \ket{X},
\end{equation}
where $\prod_{v\in X} {(c_v^\dagger)}^{N^f(\{\sigma_{a\in v}^z\})}$ creates fermions on the vertices along the string $X$ according to the fusion rule (a fermion is created on vertex $v$ if the corresponding $N^f(\{\sigma_{a\in v}^z\})=1$). $\sigma(X)=\pm 1$ is a sign factor determined by the associativity relations and the ordering of the fermion operators on $X$,
and Fig.\ref{bending} (d) implies that each closed loop contributes a factor $\alpha=\pm i$ for the ground state wavefunction.

The fermionic toric code model also has four-fold ground state degeneracy on a torus, labeled by the even/odd number of global strings wrapping around in both directions. Such a result is consistent with the Chern-Simons TQFT with $K$-matrix $K^{fTC}_{IJ}$ and $K^{\emph{fTC}}_{IJ}$ since $|\det[K^{fTC}_{IJ}]|=|\det[K^{\overline{fTC}}_{IJ}]|=4$.

Nevertheless, the most important property of the theory, is the braiding statistic of its quasiparticle excitations, which makes it different from the toric code model. It is not a surprise that the fermionic toric code model also has four different types of topological quasiparticles due to its four fold degeneracy on torus. In the following, we will compute the self-braiding $T$ matrix and mutual-braiding $S$ matrix for these quasiparticles.

\section{Algebraic model of anyons }
Algebraic models for anyons in fermion systems are more complicated because braidings of anyons are not well-defined due to the $-1$ ambiguity resulted from attaching a fermion to each anyon.  One solution is to go to a covering modular category---a spin modular category \cite{RW2008}. Physical, the covering spin modular category can be realized by ¡°gauging¡± the fermion parity, that is, couple a $\Z_2$(associated with fermion parity conservation) dynamical gauge field to a fermion system. For the fermionic toric code, we can choose the double of $\Z_4$ modular category as the cover theory.  If we label the $4$ anyons as $\{0,1,2,3\}$, we can choose $\psi_{e}=0\times 2$ as the fermion.  As is known, the Hilbert space for the torus $T^2$ associated to the covering TQFT then decomposes into a direct sum indexed by the different spin structures.  The degeneracy on the torus associated to the double of $\Z_4$ is $16$-fold.  The $4$ spin structures of the torus can be understood as the $4$ boundary conditions for the two cycles of the torus: $AP,PA, AA$ and $PP$, where $P$ is for periodic and $A$ for anti-periodic.  The direct summand for each spin structure is $4$, which gives rise to another method to find the degeneracy on the torus.  In principle, we need to choose a boundary condition in order for $H_{fTC}$ to be well-defined.  But our choice of the branching structure imposes the $PP$ boundary condition on the torus.  For the $PP$ boundary condition, both $S,T$ matrices are well-defined.

\begin{figure}[t]
\begin{center}
\includegraphics[width=7.5cm]{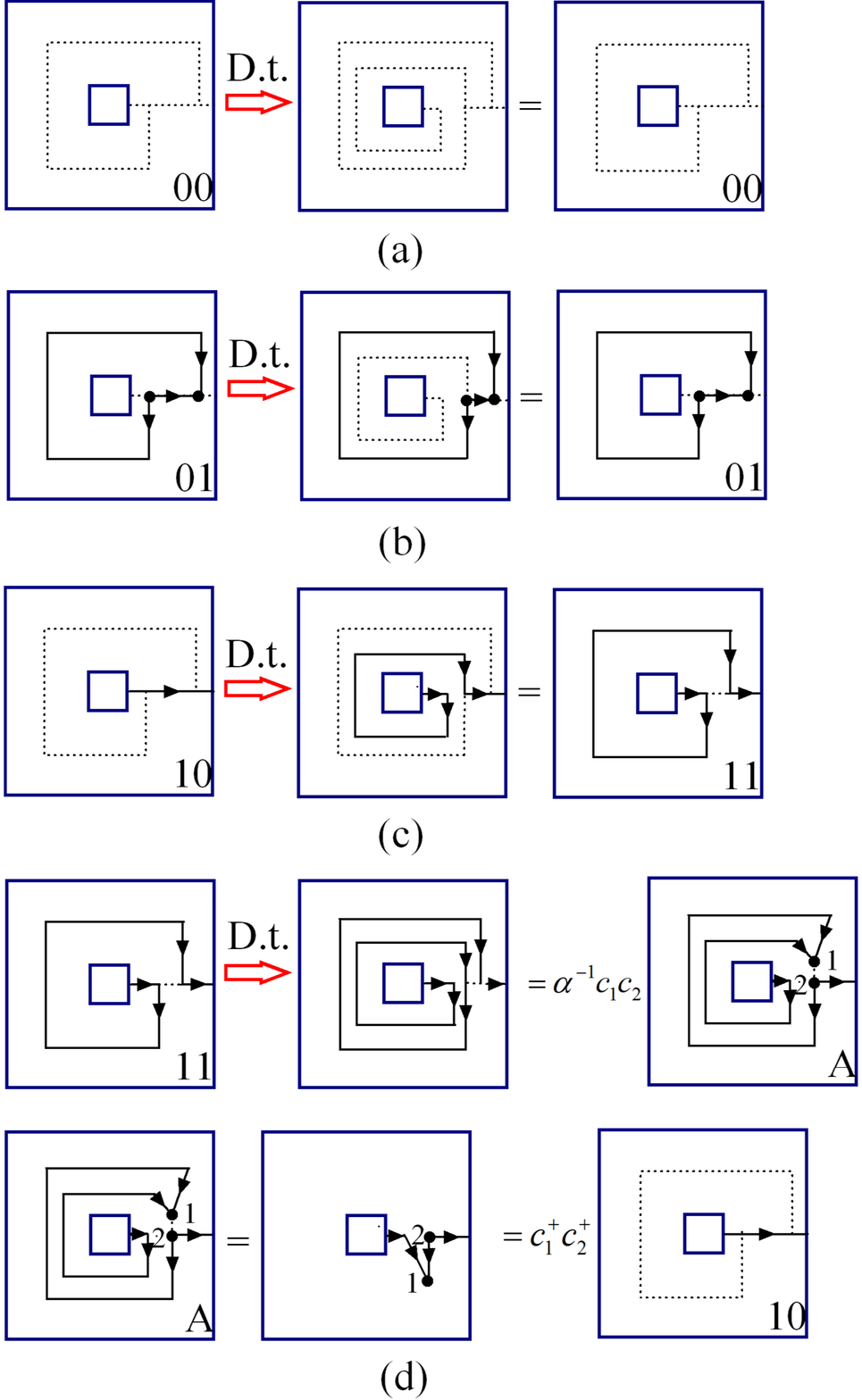}
\caption{(color online) By applying the Dehn twist and making use of the fermionic associativity relations, we are able to compute the $T$-matrix for the fermionic toric code model.
}\label{T}
\end{center}
\end{figure}

The most important data for a UMC are the modular $S$ matrix and the $T$ matrix of topological spins.
We can compute them in three ways: from the $K$-matrix formalism, the covering spin modular theory, and directly performing the $\frac{\pi}{2}$-rotation and Dehn twist on the torus.

From the $K$-matrix formalism, the $4$ anyons are labeled by vectors $v_0=(0,0),  v_1=(\frac{1}{2},0), v_2=(\frac{1}{4}, \frac{1}{2})$ and $v_3=(-\frac{1}{4}, \frac{1}{2})$.  Using the formulas $s_{ij}=e^{2\pi i <v_i|K|v_j>}$ and $\theta_i=e^{\pi i <v_i|K|v_i>}$, we obtain the $S$ and $T$ matrices:
\begin{equation}
 S=
 \frac{1}{2}\left(
   \begin{array}{cccc}
     1 & 1 & 1 & 1 \\
     1 & 1 & -1 & -1 \\
     1 & -1 & -i & i \\
     1 & -1 & i & -i \\
   \end{array}
 \right)
\end{equation}
and
\begin{equation}
 (e^{i\theta})=(1,1,e^{i\frac{3\pi}{4}},e^{i\frac{7\pi}{4}}).
\end{equation}
The same result can be obtained from the covering spin modular theory.

Now we directly compute them from our lattice model.  To compute the $T$ matrix for the fermionic toric code model, we first construct the four degenerate ground states (labeled by $|00\rangle,|01\rangle,|10\rangle$ and $|11\rangle$, representing even/odd number of global strings along both directions) on a torus.  As shown in Fig.\ref{T}, the inner square and the outer square are identified
to form a torus. Then we apply the Dehn twist\cite{Gu2010,Fangzhou13} and the associativity relation to compute the induced unitary transformation. We note that in a fixed point model, it is sufficient to study the degenerate states on a torus which only contains two sites with a fixed branching structure. By definition, the unitary transformation induced by the Denh twist is the $T$ matrix that we are looking for. As shown in Fig.\ref{T}, the states $|00\rangle$ and $|01\rangle$ remain unchanged, while $|10\rangle$ transforms to $|11\rangle$ and $|11\rangle$ transforms to $-\alpha^{-1}|10\rangle$(since $\alpha^{-1}c_1c_2|A\rangle=\alpha^{-1}c_1c_2c_1^\dagger c_2^\dagger |10\rangle=-\alpha^{-1}|10\rangle$).  In general, the numerical values of $T$ matrices are gauge dependent, but due to the special gauge choice above, these values here and below are fixed.

\begin{figure}[t]
\begin{center}
\vskip -0.5cm
\includegraphics[width=8cm]{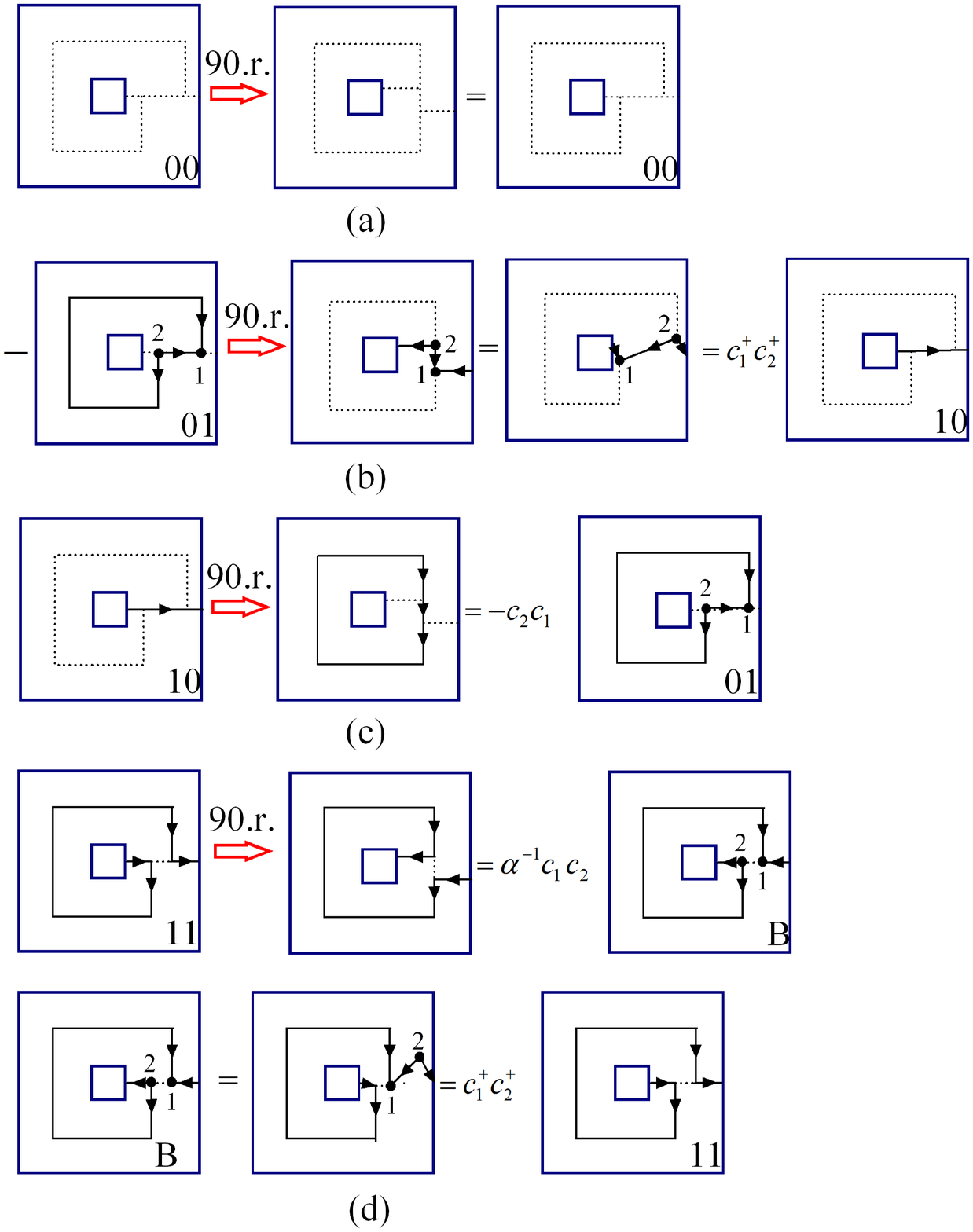}
\vskip -0.5cm
\caption{(color online) By applying the $90$-degree rotation and making use of the associativity relations, we are able to compute the $S$-matrix for the fermionic toric code model.
}\label{S}
\end{center}
\end{figure}

Similarly, by applying the $\frac{\pi}{2}$-rotation\cite{Fangzhou13}, we are able to derive the $S$ matrix, as shown in Fig.\ref{S}. However, there is a tricky issue in Fig.\ref{T} (b). We note that the minus sign on the left side is crucial since when we rotate two fermions by $\frac{\pi}{2}$, a $\pi/4$ phase factor will be induced for each fermion (the topological spin for a fermion is $\pi$) and a $\pi/2$ phase will be induced from the fermion exchange statistics. As a result, a total phase $\pi$ which corresponds to the minus sign on the left side must be taken into account. In conclusion, Fig.\ref{T} and Fig.\ref{S} imply the following $T$ and $S$ matrices:

\begin{equation}
 T=\left(
   \begin{array}{cccc}
     1 & 0 & 0 & 0 \\
     0 & 1 & 0 & 0 \\
     0 & 0 & 0 & 1 \\
     0 & 0 & -\alpha^{-1} & 0 \\
   \end{array}
 \right);
\end{equation}
and
\begin{equation}
  S=\left(
   \begin{array}{cccc}
     1 & 0 & 0 & 0 \\
     0 & 0 & - c_1^\dagger c_2^\dagger & 0 \\
     0 & - c_2c_1 & 0 & 0 \\
     0 & 0 & 0& -\alpha^{-1}  \\
   \end{array}
 \right)
\end{equation}
Interestingly, some of the elements in the $S$ matrix contain fermion creation/annihilation operators due to the different fermion parity that $|01\rangle$ and $|10\rangle$ have. To fix this problem, we can represent $T$ and $S$ matrices in the larger Hilbert space with $|00\rangle,-c_2c_1|01\rangle,|10\rangle$ and $|11\rangle$ as a basis. We finally obtain:
\begin{equation}
 T=\left(
   \begin{array}{cccc}
     1 & 0 & 0 & 0 \\
     0 & 1 & 0 & 0 \\
     0 & 0 & 0 & 1 \\
     0 & 0 & -\alpha^{-1} & 0 \\
   \end{array}
 \right);
\end{equation}
and
\begin{equation}
  S=\left(
   \begin{array}{cccc}
     1 & 0 & 0 & 0 \\
     0 & 0 & 1 & 0 \\
     0 & 1 & 0 & 0 \\
     0 & 0 & 0 & -\alpha^{-1}  \\
   \end{array}
 \right)
\end{equation}
After we diagonalize the $T$-matrix and take the numerical value of $\alpha=\pm i$, we obtain the statistic angle of the four type quasiparticles:

\begin{equation}
 (e^{i\theta})=(1,1,e^{i\frac{\pi}{4}},e^{i\frac{5\pi}{4}}) \text{  or  } (1,1,e^{i\frac{3\pi}{4}},e^{i\frac{7\pi}{4}})
\end{equation}
and in such a new basis which diagonalizes the $T$-matrix, we have:
\begin{equation}
 S=\frac{1}{2}\left(
   \begin{array}{cccc}
     1 & 1 & 1 & 1 \\
     1 & 1 & -1 & -1 \\
     1 & -1 & i & -i \\
     1 & -1 & -i & i \\
   \end{array}
 \right) \text{  or  }
 \frac{1}{2}\left(
   \begin{array}{cccc}
     1 & 1 & 1 & 1 \\
     1 & 1 & -1 & -1 \\
     1 & -1 & -i & i \\
     1 & -1 & i & -i \\
   \end{array}
 \right)
\end{equation}

Even though only $e^{i2\theta}$ is well defined for a fermion system \cite{Wrig,KW9327,FNSWW04}, the above results for $\alpha=-i$ are consistent with the $e^{i\theta}$ and the $S$ matrix described by $K^{fToric}$ above.  The value  $\alpha=i$ corresponds to $K^{\overline{fTC}}_{IJ}$.

\section{Conclusions}
In conclusion, we have constructed an exactly solvable lattice model to realize
a topologically ordered phase that cannot be realized in a local boson/spin
system.

From physical point of view, the local density matrices of interacting fermion systems always have a block diagonalized structure consisting of fermion parity even and odd sectors, that's why we need to introduce a $\Z_2$-graded structure for the fusion rules to enrich the low energy subspace on vertices\cite{CGW1038,Gu2010}(in general, the fermion can be put on links as well\cite{Gu2010}, however, the basic constructions proposed in this paper will not change), and the corresponding \emph{local} associativity relations that describe transition amplitudes between two admissible configurations differing \emph{locally} in a fixed point ground state wavefunction are also enriched over such a $\Z_2$-graded structure. Therefore, the consistent conditions for associativity relations can be defined in the enlarged Hilbert space with mixing of different fermion parities, which typically supports new solutions when a fermion loop is evolved in consistent conditions(this is a unique feature for fermion systems where fermions are viewed as identity particles). In the quantum information language, topological order in interacting fermion systems can be classified by inequivalent classes under \emph{fermionic} finite depth local unitary transformations, and the $\Z_2$-graded structure emerges naturally from the fermionic generalization of support space for a fixed point wavefunction\cite{Gu2010}.

From mathematical point of view, the key ingredient is to use Grassman number valued $6j$ symbols to
solve the pentagon equations.  These $6j$ symbols allow us to write down the
fermionic fixed point wavefunction and compute the topological data for the
theory. This general method to construct exactly solvable lattice models for
topologically ordered phases for fermion systems will be presented in Ref
\cite{Gu2010}.

Research at Perimeter Institute is supported by the Government of Canada
through Industry Canada and by the Province of Ontario through the Ministry of
Research and Innovation.  XGW is also
supported by NSF Grant No.  DMR-1005541, NSFC 11074140, and NSFC 11274192.

\bibliography{wencross,mybib,all,publst}
\appendix
\begin{widetext}
\section{The Fermionic Pentagon Equations}

In Fig.\ref{fermionPentagon}, we list all the nontrivial Pentagon relations that the fermionic associativity relations must satisfy.

\begin{figure}[h]
\begin{center}
\includegraphics[width=13cm]{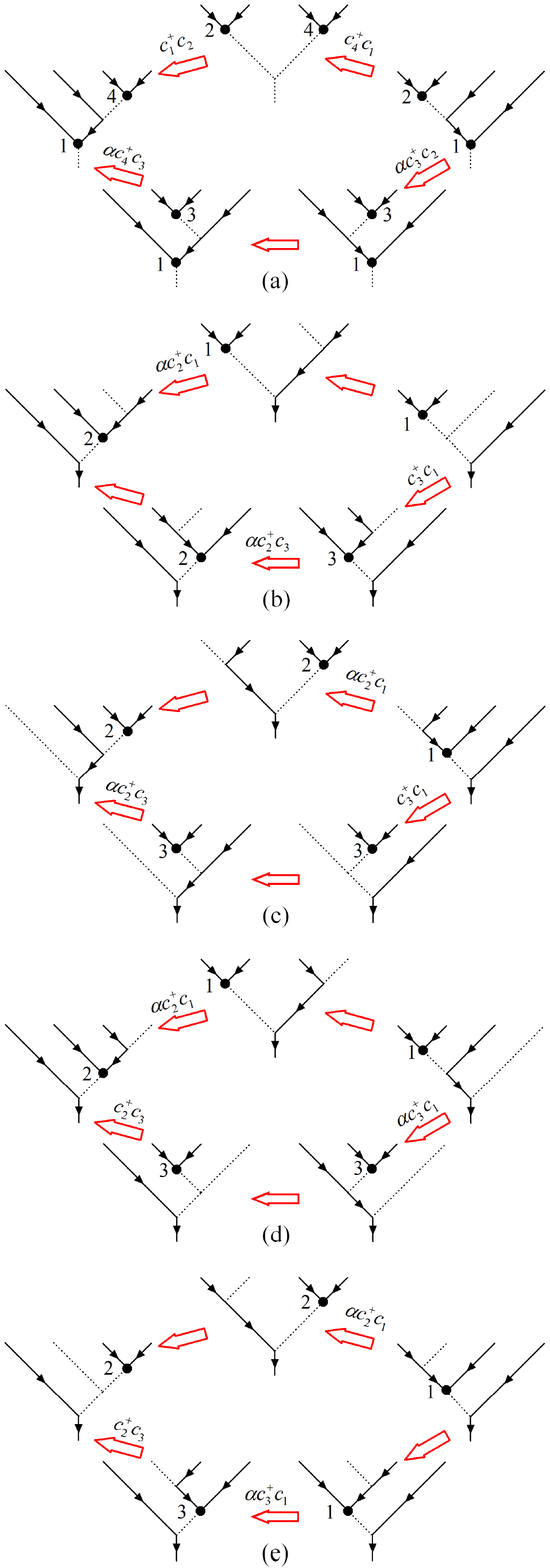}
\caption{The only five nontrivial fermionic pentagon equations that involve the nontrivial associativity relation in Fig.\ref{fermionfusion} (a), and only the first relation (a) gives rise to a constraint on the numerical values of $\alpha$. The other cases which only change the shape/length of a string or the position of a fermion automatically satisfy the Pentagon equations.
}\label{fermionPentagon}
\end{center}
\end{figure}

\end{widetext}
\end{document}